\def\e{{\rm e}}
\def\del{\partial}
\def\half{{1\over2}}

\def\abs#1{{\left|{#1}\right|}}
\def\vev#1{\langle #1 \rangle}

\def\del{\partial}
\def\dslash{\del\kern-0.55em\raise 0.14ex\hbox{/}}

\def\rough#1{\raise.3ex\hbox{$#1$\kern-.75em\lower1ex\hbox{$\sim$}}}

\newcommand{\PRD}[3]{{\it Phys. Rev.} {\bf D{#1}} (19{#3}) {#2}}
\newcommand{\PRDM}[3]{{\it Phys. Rev.} {\bf D{#1}} (20{#3}) {#2}}

\newcommand{\NPB}[3]{{\it Nucl. Phys.} {\bf B{#1}} {#2} (19{#3})}
\newcommand{\NPBM}[3]{{\it Nucl. Phys.} {\bf B{#1}} (20{#2}) {#3}}
\newcommand{\PLB}[3]{{\it Phys. Lett.} {\bf {#1}B} (19{#3}) {#2}}
\newcommand{\PLBM}[3]{{\it Phys. Lett.} {\bf {#1}B} (20{#3}) {#2}}

\newcommand{\ANN}[3]{{\it Ann. Phys. (N.Y.)} {\bf {#1}}, {#2} (19{#3})}

\newcommand{\MPL}[3]{{\it Mod. Phys. Lett.} {\bf A{#1}} (19{#3}) {#2}}
\newcommand{\MPLM}[3]{{\it Mod. Phys. Lett.} {\bf A{#1}} (20{#3}) {#2}}

\newcommand{\jhep}[3]{{\it J. High Energy Phys.}{\bf {#1}}, {#2} (20{#3})}
\newcommand{\hepth}[1]{{\tt [hep-th/#1]}}

\newcommand{\hmu}{\hat\mu}

\documentclass[12pt]{article}
\usepackage{graphicx}
\begin{document}
\baselineskip=18pt
\begin{titlepage}
\begin{flushright}
OU-HET-536/2005
\end{flushright}
\begin{center}{\Large\bf Effective Potential 
of Super Yang-Mills Theory on 
$M^4\times S^1$ \\ and \\ related topics}
\footnote{Based on a talk given at Planck '05, 23-28 May, 2005,
ICTP, Trieste}
\end{center}
\vspace{0.5cm}
\begin{center}
%
%
Kazunori {Takenaga}
\footnote{E-mail: takenaga@het.phys.sci.osaka-u.ac.jp}
%
\end{center}
\vspace{0.2cm}
\begin{center}
%
{\it Department of Physics, Osaka University, 
Toyonaka, Osaka 560-0043, Japan}
\end{center}
\vspace{1cm}
\begin{abstract}
We study the gauge symmetry breaking of 
an ${\cal N}=1$ supersymmetric Yang-Mills theory 
defined on $M^4\times S^1$, taking correctly account of 
the vacuum expectation values for the 
adjoint scalar field $\vev{\Sigma}$ in vector 
multiplet in addition to the Wilson line phases $\vev{A_y}$. 
We evaluate the one-loop effective potential and obtain the
vacuum configuration, for which an $SU(N)$ gauge symmetry is not 
broken. In case of an orbifold $S^1/Z_2$, under appropriate
orbifolding boundary conditions, two Higgs doublets are embedded
in the zero modes, $A_y^{(0)}$ and $\Sigma^{(0)}$. We point out that 
the tree-level scalar potential resulted 
from the covariant derivative for the adjoint scalar field
is identical to the $D$-term of the MSSM.
We also briefly mention the mass spectra of the gauge and 
Higgs sector in the theory.
\end{abstract}
\end{titlepage}
\section{Introduction}
Gauge theory in higher dimensions is able to provide new 
approaches to the long standing problems in high energy 
physics. One of the interesting ideas in the higher dimensional 
gauge theory is the gauge-Higgs 
unification \cite{gaugehiggs}, where scalar 
fields are unified into the higher dimensional gauge 
field as extra components. The extra component, in fact, behaves 
as the scalar field at low energies.
\par
The zero mode of the extra component of the gauge field becomes
dynamical variable to induce the vacuum expectation values,
reflecting the topology of the extra dimension.
The vacuum expectation values are closely related to the Wilson
line phase, and one usually evaluates the effective potential
for the phases and the vacuum expectation
values is determined dynamically. The zero mode is massless at
the tree-level, but it acquires mass term at quantum level. The mass
term is obtained from the effective potential. 
\par
If we can identify the Higgs scalar in the standard model as the extra
component of the gauge field, namely, its zero mode, the
arbitrariness of the Higgs sector in the standard model is 
partially solved.
The Higgs self interactions appears from the higher dimensional gauge
coupling, and more interestingly, the gauge hierarchy problem is
resolved. This is because the Higgs mass is
generated through quantum corrections to be finite from the effective
potential for the Wilson line phases. The dynamics of the 
Wilson line phases, the Hosotani mechanism \cite{hosotani} plays 
the crucial role.  
\par
We study an ${\cal N}=1$ vector multiplet on $M^4\times S^1$ and
obtain the one-loop effective potential, taking account of the
vacuum expectation values for the adjoint scalar field, which has
been overlooked in the past, in addition to the Wilson line
phases \cite{hty}. We also study the case of two Higgs doublets by
proceeding to an orbifold $S^1/Z_2$ and point out that the scalar
potential is identical to that of the MSSM with
$g_Y=\sqrt{3}g_2$. We briefly mention the mass spectra of the
gauge and Higgs sector. 
\section{Effective potential $V_{eff}(\vev{A_y}, \vev{\Sigma})$}
Let us consider an ${\cal N}=1$ vector 
multiplet $(A_{\hmu}, \Sigma,\lambda_D)$ on $M^4\times S^1$.
As is well known, reflecting the topology of $S^1$, the order parameter
for the gauge symmetry breaking is given by the zero mode of the extra
component of the gauge potential $A_y^{(0)}$. In addition to it, one
should not overlook the adjoint scalar field $\Sigma$, which also
carries the colour indices. Hence, there are two kinds of the order
parameters for the gauge symmetry breaking,
\begin{equation}
\vev{A_y^{(0)}}=\mbox{Wilson~line~phases}
\qquad
\mbox{and}
\qquad
\vev{\Sigma}\in \mbox{Adj.~representation}. 
\label{shiki1}
\end{equation}
In order to study the vacuum structure of the theory, one needs to take
both into account. We expand fields around the VEV's,
\begin{equation} 
A_{\hmu}=\vev{A_{\hmu}}\delta_{\hmu y}+{\bar A}_{\hmu},\qquad 
\Sigma=\vev{\Sigma}+{\bar \Sigma}.
\label{shiki2}
\end{equation}
Thanks to $\vev{\Sigma}$, the tree-level potential arises from
the covariant derivative for $\Sigma$,
\begin{equation}
V_{tree}=-g^2\mbox{tr}[\vev{A_y},~\vev{\Sigma}]^2.
\label{shiki3}
\end{equation}
By utilizing the $SU(N)$ degrees of freedom, we can diagonalize
$\vev{A_y}$ as
\begin{equation}
gL\vev{A_y}=(\theta_1, \theta_2,\cdots,\theta_N)\quad \mbox{with}\quad
\sum_{i=1}^N\theta_i=0.
\label{shiki4}
\end{equation}
It is natural to expect that the vacuum configuration satisfies the
flatness condition, $[\vev{A_y},~\vev{\Sigma}]=0$, so that
we can parametrize $\vev{\Sigma}$ as
\begin{equation}
\vev{\Sigma}=\mbox{diag.}(\sigma_1,\sigma_2,\cdots,\sigma_N)
\quad \mbox{with}\quad
\sum_{i=1}^N\sigma_i=0.
\label{shiki5}
\end{equation}
\par
The contribution to the VEV's from bosons and fermions cancels due
to the supersymmetry, so that the effective potential vanishes. One
needs to break the supersymmetry in order to obtain the 
nonvanishing effective potential. There is a simple framework 
to break supersymmetry in studying the dynamics of the Wilson 
line phases. That is the Scherk-Schwarz (SS) 
mechanism \cite{ss}, by which 
the boundary condition of $\lambda_D$ in the vector multiplet 
for the $S^1$ direction is twisted by an amount of $\beta$,
\begin{equation}
\lambda_D(y+L)=\e^{2\pi \beta}\lambda_D(y).
\label{shiki6}
\end{equation}
The other fields satisfy the periodic boundary condition. The
nontrivial
values for $\beta$ explicitly breaks supersymmetry.
\par
One can also introduce the gauge invariant mass $M$ for
$\lambda_D$ to break supersymmetry. In this case, however, one 
should notice that the $\sigma_i$-dependent divergent terms 
like $M\vev{\Sigma}^2\Lambda$ appear
to spoil the desirable nature of the ultraviolet insensitivity
for the effective potential. One can
formally remove the divergent by subtracting the $n=0$ mode in the
KK mode summation, which
corresponds to the contribution from $L\rightarrow \infty$.  
Here we consider the SS mechanism alone, so that the effective
potential is independent of the ultraviolet cutoff.
\par
By the straightforward calculations \cite{pomarol}, we arrive at
\begin{eqnarray}
V_{eff}(\sigma, \theta)
&=&{{-4\times 2}\over{(2\pi)^{5\over 2}}}
\sum_{i, j=1}^N\sum_{n=1}^{\infty}
\left({{g^2\sigma_{ij}^2}\over {n^2L^2}}\right)^{5\over 4}
K_{5\over 2}\left(\sqrt{(g\sigma_{ij}nL)^2}\right)
\left[1- \cos(2\pi n\beta)\right]
\nonumber\\
&\times & 2\cos[n(\theta_i-\theta_j)],\qquad  \sigma_{ij}\equiv 
\sigma_i-\sigma_j,
\label{shiki7}
\end{eqnarray}
where the modified Bessel function is expressed as
\begin{equation}
K_{5\over 2}(y)=\left({\pi\over {2y}}\right)^{\half}
\left(1 + {3\over y} +{3\over y^2} \right)\e^{-y}.
\label{shiki8}
\end{equation}
\par
The Boltzmann like suppression factor $\e^{-gn\sigma_{ij}L}$ is 
understood from the similarity of the effective potential 
as that at finite temperature. Particles with 
smaller wavelengths 
than the inverse temperature $(\sim L)$ have the Boltzmann 
suppressed distribution in the system. It has been known that the 
factor is important for gauge symmetry breaking through 
the Wilson line phases \cite{takenaga}\cite{yamashita}.
\par
Noting that 
\begin{equation}
0\leq K_{5\over 2}\left(\sqrt{(g\sigma_{ij}nL)^2}\right)
[1-\cos(2\pi n\beta)]\quad \forall \sigma_{ij}, \beta,
\label{shiki9}
\end{equation}
we immediately see that the vacuum configuratio is given by
\begin{equation}
\left(\theta_i, \sigma_i\right)=\left( {{2\pi k}\over N}, 0 \right)\quad
k=0,1,\cdots, N-1.
\label{shiki10}
\end{equation}
Since the configuration for $\theta_i$
is the center of $SU(N)$, the $SU(N)$ gauge symmetry is unbroken for the
vacuum configuration. 
\par
The zero modes $A_y^{(0)a=3}$ and $\Sigma^{a=3}$ become massive at
one-loop level, and their masses, for example $SU(2)$ case, are 
obtained by the mass matrix,
\begin{equation}
{\cal M}_{A_y, \Sigma}^2=\left(
\begin{array}{cc}
{{\del^2 V_{eff}}\over {\del\theta^2}} & 
{{\del^2 V_{eff}}\over{\del\theta \del\sigma}} 
\\[0.3cm]
{{\del^2 V_{eff}}\over{\del\theta \del\sigma}}&
{{\del^2 V_{eff}}\over {\del\sigma^2}} 
\end{array}
\right)_{\theta=0, \sigma=0}.
\label{shiki11}
\end{equation}
The off-diagonal elements vanishes for the vacuum configuration, and we
obtain that
\begin{equation}
m_{A_y^{(0)a=3}}^2=3m_{\Sigma^{a=3}}^2=\left(g_2 \over L\right)^2 
{6 \over \pi^2}g(\beta),
\quad 
g(\beta)=\sum_{n=1}^{\infty}{1\over n^3}\left[1-\cos(2\pi n\beta)\right],
\label{shiki12}
\end{equation}
where $g_2\equiv g/\sqrt{2\pi R}$.
\section{An orbifold $S^1/Z_2$ case}
Let us proceed to an orbifold $S^1/Z_2$ case. There are two orbifold
fixed points, $y=0, \pi R$. We must specify the boundary condition of
the field at the fixed points in addition to the $S^1$ direction.
By choosing the appropriate orbifolding boundary conditions, it is
possible to embed the two Higgs doublets $\chi_i (i=1,2)$ into
the zero modes $A_y^{(0)},\Sigma^{(0)}$, 
\begin{eqnarray}
A_y^{(0)}&=&\half\left(
\begin{array}{cc|c}
& &A_y^{4}-iA_y^{5} \\ 
& &A_y^{6}-iA_y^{7} \\ \cline{1-3}
\mbox{c.c.}&\mbox{c.c.} & 
\end{array}
\right)
\equiv
{1\over{\sqrt{2\pi R}}}
\left(
\begin{array}{cc|c}
& & {\chi_1\over \sqrt{2}}\\ 
& & \\ \cline{1-3}
{\chi_1^{\dagger}\over\sqrt{2}}& & 
\end{array}
\right),\label{shiki13}
\\
\Sigma^{(0)}&=&\half\left(
\begin{array}{cc|c}
& &\Sigma^{4}-i\Sigma^{5} \\ 
& &\Sigma^{6}-i\Sigma^{7} \\ \cline{1-3}
\mbox{c.c.}&\mbox{c.c.} & 
\end{array}
\right)
\equiv 
{1\over{\sqrt{2\pi R}}}
\left(
\begin{array}{cc|c}
& & {\chi_2\over\sqrt{2}}\\ 
& & \\ \cline{1-3}
{\chi_2^{\dagger}\over\sqrt{2}}& & 
\end{array}
\right).
\label{shiki14}
\end{eqnarray}
Then, the tree level potential is written, in terms 
of \footnote{The complex scalar field in the ${\cal N}=2$
vector multiplet in four dimensions is given by $\phi=A_y+i\Sigma$.}
\begin{equation}
\Phi_1={1\over\sqrt{2}}(\chi_1 -i\chi_2),\quad 
\Phi_2={1\over\sqrt{2}}(\chi_1 +i\chi_2),
\label{shiki15}
\end{equation}
as
\begin{eqnarray}
V_{tree}&=&-g^2\mbox{tr}[A_y^{(0)},~\Sigma^{(0)}]^2
\nonumber\\
&=&{g_2^2\over 2}
\left(
(\Phi_1^{\dagger}\Phi_1)^2+(\Phi_2^{\dagger}\Phi_2)^2
-(\Phi_1^{\dagger}\Phi_1)
(\Phi_2^{\dagger}\Phi_2)-\abs{\Phi_1^{\dagger}\Phi_2}^2
\right).
\label{shiki16}
\end{eqnarray}
This is identical with the $D$-terms of the MSSM with
$g_Y=\sqrt{3}g_4$. The relation means that the Weinberg angle is too 
large, $\sin^2\theta_w=3/4$. The potential (\ref{shiki16}) has 
the flat direction $\Phi_1=\Phi_2 (\mbox{mod~phase})$. If we 
suppose that the vacuum configuration is in the flat 
direction, this implies 
that $\tan\beta\equiv \abs{v_2}/\abs{v_1}=1$,
where $\Phi_{1(2)}\equiv {v_{1(2)}\over\sqrt{2}}{0\choose 1}$.
\par
The charged and heavier neutral (CP-even) Higgses are 
massive at the tree level due to the quartic coupling (\ref{shiki16}),
\begin{equation}
M_{H^{\pm}}^2={g_2^2\over 4}v^2 (=M_W^2),\quad
M_H^2={{g_2^2+g_Y^2}\over 4}v^2=g_2^2v^2(=M_Z^2). 
\end{equation}
In terms of the original parametrizations, we have 
\begin{equation}
v^2(=\abs{v_1}^2+\abs{v_2}^2)
={1\over g_2^2}\left(({a\over R})^2+(gp)^2\right),
\end{equation}
where $\vev{A_y^{(6)}}=a/gR, \vev{\Sigma^{(6)}}=p$. Let us note
that there is no $g_2$-dependence in the tree-level mass
spectra. On the other hand, the 
lighter (CP-even) and CP-odd Higgses become massive 
at one-loop level, which are calculated
by the effective potential $V_{eff}(\theta, \sigma)$, whose 
form is similar to (\ref{shiki7}). Their magnitude is
order of $O(g_2^2)$ because they are generated at one-loop level. 
We also note that the corresponding SUSY breaking mass 
parameters $m_i^2 (i=1,2,3)$ are also obtained from 
the effective potential.
\par
Let us notice that one can obtain the same quartic coupling as
(\ref{shiki16}) by starting with the six dimensional pure Yang-Mills
theory compactified on $M^4\times T^2/Z_2$. In this case, the two Higgs
doublets are embedded in the zero modes $A_{y, z}^{(0)}$. It is easy to
see that
\begin{equation}
V_{tree}=-g^2\mbox{tr}[A_y^{(0)},~A_z^{(0)}]^2
\label{shiki17}
\end{equation}
gives the same form as (\ref{shiki16}) under the linear 
combinations \footnote{The form of the quartic coupling depends
on the base one chooses.},
\begin{equation}
\Phi_1={1\over\sqrt{2}}(A_y^{(0)}-iA_z^{(0)}),\quad
\Phi_2={1\over\sqrt{2}}(A_y^{(0)}+iA_z^{(0)}).
\label{shiki18}
\end{equation}
Both $A_y^{(0)}$ and $A_z^{(0)}$, however, correspond to 
the Wilson line phases, which
is very different from the five dimensional case. Accordingly, the form of the
effective potential is also quite different from that of the five
dimensional case. There appears no Boltzmann like suppression factor
in the effective potential for the present case. It is
interesting to note that the nonsupersymmetric theory
shares the feature of supersymmetric gauge theory through the dimensional 
reduction \footnote{It has been pointed out that the
supersymmetric quantum
mechanical structure is always hidden in higher dimensional
gauge theories \cite{sakamoto}.}.   
\section{Conclusions}
We have considered the ${\cal N}=1$ vector multiplet on $M^4\times
S^1$. In order to study the vacuum structure, we have 
taken into account the vacuum expectation values for 
the adjoint scalar field, which has been overlooked in the past,
in addition to the 
Wilson line phases and obtained the effective potential. 
The effect of the VEV for the adjoint scalar appears as the Boltzmann
like suppression factor in the effective potential. 
We have found that the configuration that
minimizes the effective potential does not break 
the gauge symmetry (\ref{shiki10}). 
\par
If we introduce hyper multiplets, we expect the vanishing 
VEV for $\Sigma$, but nontrivial values for $\theta$, which 
is the signal for the gauge symmetry breaking. Let us note 
that in this case the squark field $\phi_q$ is also the 
order parameter for the gauge symmetry breaking.
\par
In case of an orbifold $S^1/Z_2$, it is possible to have two 
Higgs doublets as the zero modes of $A_y$ and $\Sigma$ under the
appropriate orbifolding boundary conditions. The tree level potential is
same as the $D$-terms of the MSSM with $g_Y=\sqrt{3}g_4$. The lighter
(CP-even) and CP-odd Higgs become massive at one-loop level. The
other Higgses are
massive at the tree-level due to the quartic couplings (\ref{shiki16}). 
The same quartic coupling is obtained from the six dimensional pure
Yang-Mills theory compactified on $M^4\times T^2/Z_2$, but in
this case the two Higgs doublets correspond to the Wilson line
phases, so that the form of the effective potential is different
from the one for five dimensions.  
\vspace{2cm}
\begin{center}
{\bf Acknowledgements}
\end{center}  
The author would thank Professors N. Haba, Y. Hosotani 
and Dr. T. Yamashita for valuable 
discussions, and he is supported in part by 
the Grant-in-Aid for Science Research, Ministry of Education, Science
and Culture, Japan, No. 17043007 and 
the $21$st Century 
COE Program at Osaka University.
\vskip 2cm

\end{document}